\documentclass{optica-article}

\journal{opticajournal} 

\articletype{Research Article}

\usepackage{float}

\hypersetup{linkcolor=urlblue,citecolor=urlblue,urlcolor=urlblue}

\begin{document}

\title{Dual-Faraday-laser-pumped cesium beam clock with $7.7 \times 10^{-13}/\sqrt{\tau}$ frequency stability}

\author{Xiaomin Qin,\authormark{1} Suyang Wei,\authormark{1} Haijun Chen,\authormark{2} Yufei Yan,\authormark{2} Qiang Wei,\authormark{3} Hangbo Shi,\authormark{1} Zhiyang Wang,\authormark{1} Zheng Xiao,\authormark{1} Zijie Liu,\authormark{1} Tiantian Shi,\authormark{4,5,*} and Jingbiao Chen\authormark{1,4,6,7}}

\address{\authormark{1}School of Electronics, Peking University, Beijing 100871, China\\
\authormark{2}National Key Laboratory of Science and Technology on Vacuum Electronics, Beijing Vacuum Electronics Research Institute, Beijing 100871, China\\
\authormark{3}Longgang Advanced Quantum Technology Institute Co., Ltd., Wenzhou 325802, China\\
\authormark{4}National Key Laboratory of Advanced Micro and Nano Manufacture Technology, School of Integrated Circuits, Peking University, Beijing 100871, China\\
\authormark{5}Beijing Advanced Innovation Center for Integrated Circuits, Beijing 100871, China\\
\authormark{6}Beijing Key Laboratory of Quantum Metrology Technology and Instruments, Beijing 100871, China\\
\authormark{7}Hefei National Laboratory, Hefei 230088, China}

\email{\authormark{*}tts@pku.edu.cn} 


\begin{abstract*}

Compact cesium beam clocks are major frequency references for deployable timing systems. However, further improvement of their short-term frequency stability is limited by the clock signal-to-noise ratio (SNR). Although two-laser optical pumping can increase the effective atomic utilization, the achievable clock SNR has long been limited by laser-induced frequency-to-amplitude noise conversion. Here, we demonstrate a compact dual-Faraday-laser-pumped (DFP) Cs beam clock enabled by a low-frequency-noise atom-referenced laser architecture. The intracavity Faraday anomalous dispersion optical filter provides inherent alignment to the Cs D$_2$ resonances, while modulation transfer spectroscopy offers suppressed frequency noise and drift. The resulting laser system supports robust turnkey operation with a Lorentzian linewidth of 2.12~kHz. The DFP Cs clock achieves a clock SNR of 46,365 in a 1-Hz bandwidth and a fractional Allan deviation of $7.7\times10^{-13}/\sqrt{\tau}$, with Hadamard deviation reaching $7.4\times10^{-15}$ at 10,000~s. This work pushes the fractional frequency stability of a compact Cs beam clock into the $10^{-13}/\sqrt{\tau}$ regime, providing a pathway toward high-performance Cs frequency references for field-deployable precision timing, navigation, and synchronization.

\end{abstract*}


\section{INTRODUCTION}

Cesium atomic clocks, based on the unperturbed ground-state hyperfine transition of $^{133}$Cs that defines the SI second, have long served as primary standards in time and frequency metrology~\cite{2019-SIsecond,2018-PTB-FOUNTAIN,2025-NIM6}. For field-deployable timing applications, including global navigation satellite systems (GNSS)~\cite{2021-GPS-GNSS,2021-gnss}, telecommunication networks~\cite{2003-mst-csclock}, and other positioning, navigation and timing (PNT) systems~\cite{2004-science-Cs,2026-NC-Cs}, compact Cs beam clocks remain attractive because of their high accuracy, low long-term drift, independence from an optical frequency comb, and mature commercial engineering~\cite{2005-metrologia-years,2005-fifty}, despite the rapid progress of high-performance portable optical clocks~\cite{2017-PRL-OPTICALCLOCK,2020-NP-OPTICALCLOCK,2026-optica-Yb} and chip-scale atomic clocks~\cite{2002-apl-chip,2018-APR-chip,2023-nc-chip}. Exemplifying this engineering legacy, the 5071A has served as a workhorse of international timekeeping for more than three decades~\cite{5071A}, establishing a benchmark for the reliability and operational robustness of commercial Cs beam clocks. Building on these proven strengths, advancing the frequency stability of compact Cs beam clocks beyond the prevailing low-$10^{-12}/\sqrt{\tau}$ regime~\cite{2020-OE-SHS,2024-PR-STT,2025-prapplied-shb,2026-prapplied-wyh} represents an important pathway toward next-generation deployable timing systems.

For optically pumped Cs beam clocks based on Ramsey excitation~\cite{1950-Ramsey}, the short-term frequency stability is mainly determined by the signal-to-noise ratio (SNR) of the Ramsey signal. Compared with the traditional single-laser pumping scheme, two-laser optical pumping can increase the atomic utilization by redistributing population into the $m_F=0$ clock state~\cite{1981-IFCS-D,1984-jdp-D}. However, this increase in signal amplitude has not been translated into a better clock SNR~\cite{1985-TIM-D,1988-TIM-D}. Since the laser fields determine both state preparation and fluorescence detection, their frequency fluctuations are converted into fluorescence fluctuations, ultimately introducing excess noise in the Ramsey signal. Unlike independent noise sources such as atomic shot noise, this laser-induced noise can scale quadratically with atomic flux and limit further SNR improvements in high-atom-flux operation of the two-laser-pumped clock, and is especially evident for lasers with high noise. As a result, the atomic utilization advantage of two-laser pumping can only be converted into clock stability improvement when the laser-induced noise is sufficiently suppressed.

In compact Cs clocks, monolithic distributed-feedback (DFB) and distributed-Bragg-reflector (DBR) laser diodes are commonly used because of their small size, low power consumption, and ease of integration~\cite{TA1000,OSA3300}. However, their MHz-level free-running linewidth, frequency noise, and slow drift fundamentally limit high-SNR operation in Cs clocks. External-cavity diode lasers (ECDLs) provide substantially narrower linewidths~\cite{1980-ecdl}, but their frequencies are determined by external-cavity modes and macroscopic frequency-selective elements such as gratings~\cite{2001-rsi-grating} or interference filters (IF)~\cite{2012-rsi-IF}. Consequently, thermal and mechanical perturbations, as well as component aging, can lead to frequency drift or mode hops, making it challenging for the frequency to remain near a specific atomic transition over extended periods.

The atom-referenced Faraday laser incorporating a Faraday anomalous dispersion optical filter (FADOF) offers an alternative approach~\cite{2011-RSI-MXY,2025-prapplied-shb}. The FADOF exploits magneto-optical rotation in Cs vapor to form a GHz-scale transmission window centered near the Cs D$_2$ resonances~\cite{2024-IEEE-WZY,2025-APL-WZY}, thereby constraining the lasing frequency to the vicinity of the atomic transitions. Compared with conventional ECDLs mentioned above, this atom-referenced Faraday laser offers two practical advantages. First, it reduces the frequency sensitivity to diode temperature and injection current variations, ensuring robust long-term operation near the target atomic transition. Second, the GHz-bandwidth frequency selection, together with a mechanically stable external-cavity structure, supports narrow-linewidth and low-noise laser operation. These features make it a promising source for compact atomic clocks and other quantum precision measurement systems.

Here, we demonstrate a dual-Faraday-laser-pumped (DFP) Cs clock based on the custom-designed atom-referenced laser scheme, in which effective suppression of laser noise enables the enhanced atomic utilization of two-laser pumping to translate into a higher Ramsey-signal SNR. For each laser, FADOF-based external-cavity feedback constrains the laser frequency to the vicinity of its targeted Cs D$_2$ transition, while modulation transfer spectroscopy (MTS) provides fine frequency stabilization. It supports robust turnkey operation and yields a Lorentzian linewidth of 2.12~kHz. The DFP Cs beam clock achieves an SNR of 46,365 in a 1-Hz bandwidth and a short-term fractional frequency stability of $7.7\times10^{-13}/\sqrt{\tau}$. With optical-power stabilization, the Hadamard deviation (HDEV) reaches $7.4\times10^{-15}$ at an averaging time of $10^4$~s. These results demonstrate a practical route for advancing the short-term frequency stability of compact Cs beam clocks into the $10^{-13}/\sqrt{\tau}$ regime, narrowing the performance gap with more complex cold-atom frequency references~\cite{crb-clock} while retaining the compactness, robustness, and deployability required for field applications.

\begin{figure}[htbp]
\centering
\includegraphics[width=12cm]{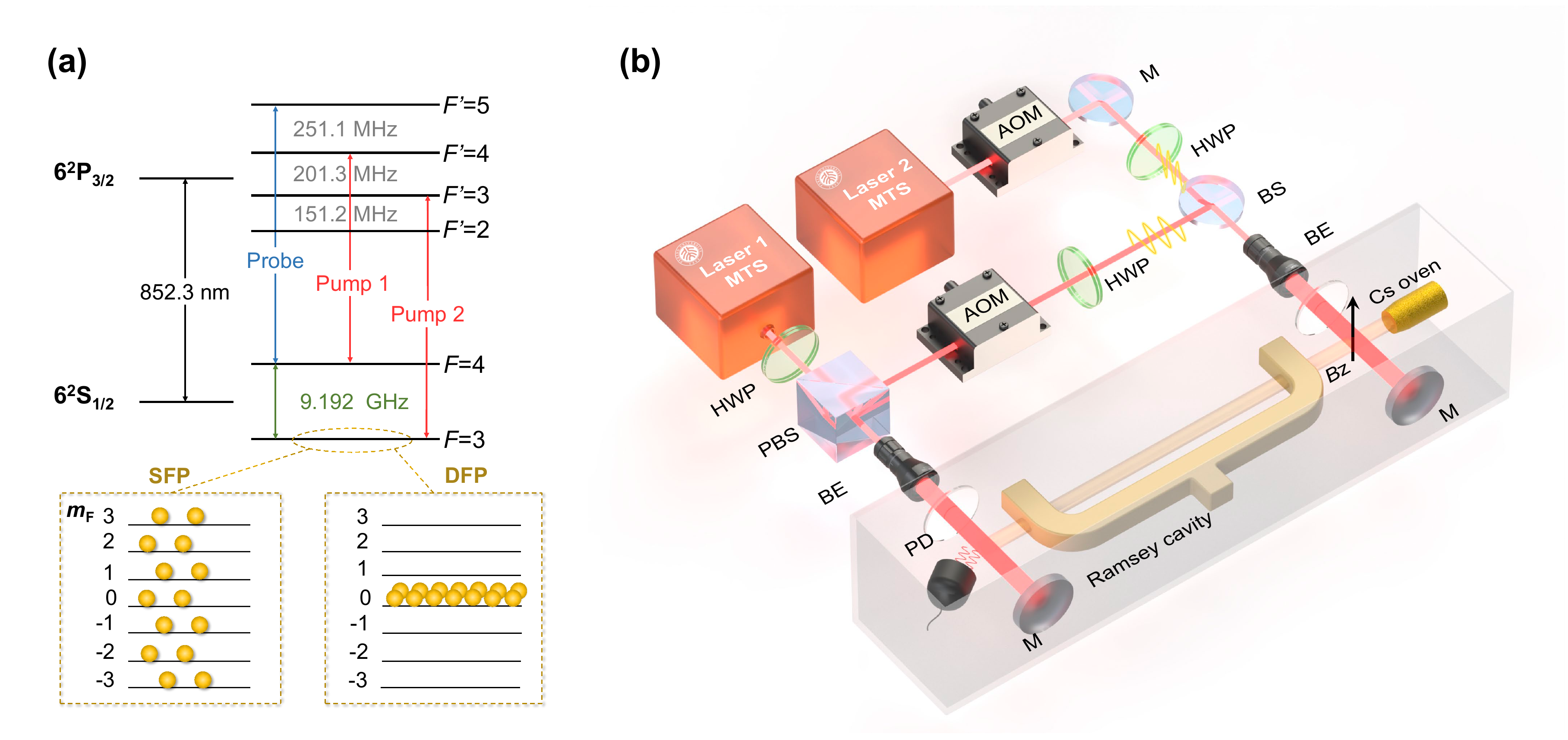}
\caption{\label{Fig.1} (a) The $^{133}$Cs D$_2$ energy-level structure and schematic diagram illustrating atom population distribution in the ground state $F=3$ under single-Faraday-laser-pumping (SFP) and dual-Faraday-laser-pumping (DFP) scheme. (b) Experimental schematic of the compact DFP Cs clock. Laser~1 and Laser~2 are atom-referenced Faraday lasers stabilized via modulation transfer spectroscopy (MTS) to the $F=4\rightarrow F'=5$ and $F=3\rightarrow F'=2$ cycling transitions, respectively. The grey region represents the sealed cesium beam tube. The black arrow represents the direction of magnetic field produced by the Helmholtz coils. Pump~1 and Pump~2 polarizations are set to $\sigma$ and $\pi$, respectively, as indicated by the yellow curves. HWP, half-wave plate; PBS, polarizing beam splitter; BE, beam expander; PD, photodetector; M, mirror; AOM, acousto-optic modulator; BS, beam splitter.}
\end{figure}

\section{EXPERIMENTAL SETUP}
\subsection{DFP Clock Configuration}

As shown in Fig.~\ref{Fig.1}, our DFP Cs clock employs a dual-Faraday-laser architecture to maximize the effective atomic flux $I_{\text{at}}$. Laser~1, frequency-stabilized to the Cs D$_2$ line $6^2\mathrm{S}_{1/2}(F=4)\rightarrow 6^2\mathrm{P}_{3/2}(F'=5)$ transition via MTS, is split by a half-wave plate (HWP) and polarizing beam splitter (PBS). The transmitted beam serves as the probe, while the reflected beam is frequency-shifted by $-251.1$~MHz via an acousto-optic modulator (AOM) to the $F=4\rightarrow F'=4$ transition, acting as Pump~1 ($\sigma$-polarized) to transfer atoms from the $F=4$ ground state into $F=3$. Laser~2, locked to the $F=3 \rightarrow F'=2$ transition, is shifted by $+151.2$~MHz to the $F=3 \rightarrow F'=3$ transition, and acts as Pump~2 ($\pi$-polarized). According to selection rules, the $|F=3,\mathrm{m_F=0}\rangle \rightarrow |F'=3,\mathrm{m_F=0}\rangle$ transition is forbidden for $\pi$-polarized light, causing atoms to accumulate in the $|F=3,\mathrm{m_F=0}\rangle$ state, as shown in Fig.~\ref{Fig.1}(a). After passing through the iris diaphragms, the pump and probe beams are expanded to a diameter of 4~mm using beam expanders (BEs), ensuring a sufficiently long light--atom interaction time.

The commercial sealed Cs beam tube used in this work consists of a Cs oven, a pumping region, a 220-mm-long Ramsey cavity, and a detection region. A uniform C-field of 60~mG is applied within the Ramsey cavity region to separate the degenerate magnetic sub-levels, with a double-layer magnetic shielding utilized to attenuate the ambient magnetic field. To determine the polarization of Pump~1 and Pump~2, three pairs of orthogonal Helmholtz coils are deployed in the unshielded pumping region to compensate the geomagnetic field and establish a well-defined magnetic quantization field along the $z$ direction. After optimization in Section~\ref{S3.1}, the polarization of Pump~1 and Pump~2 are set perpendicular and parallel to the quantization axis, respectively.

\begin{figure}[htbp]
\centering
\includegraphics[width=13cm]{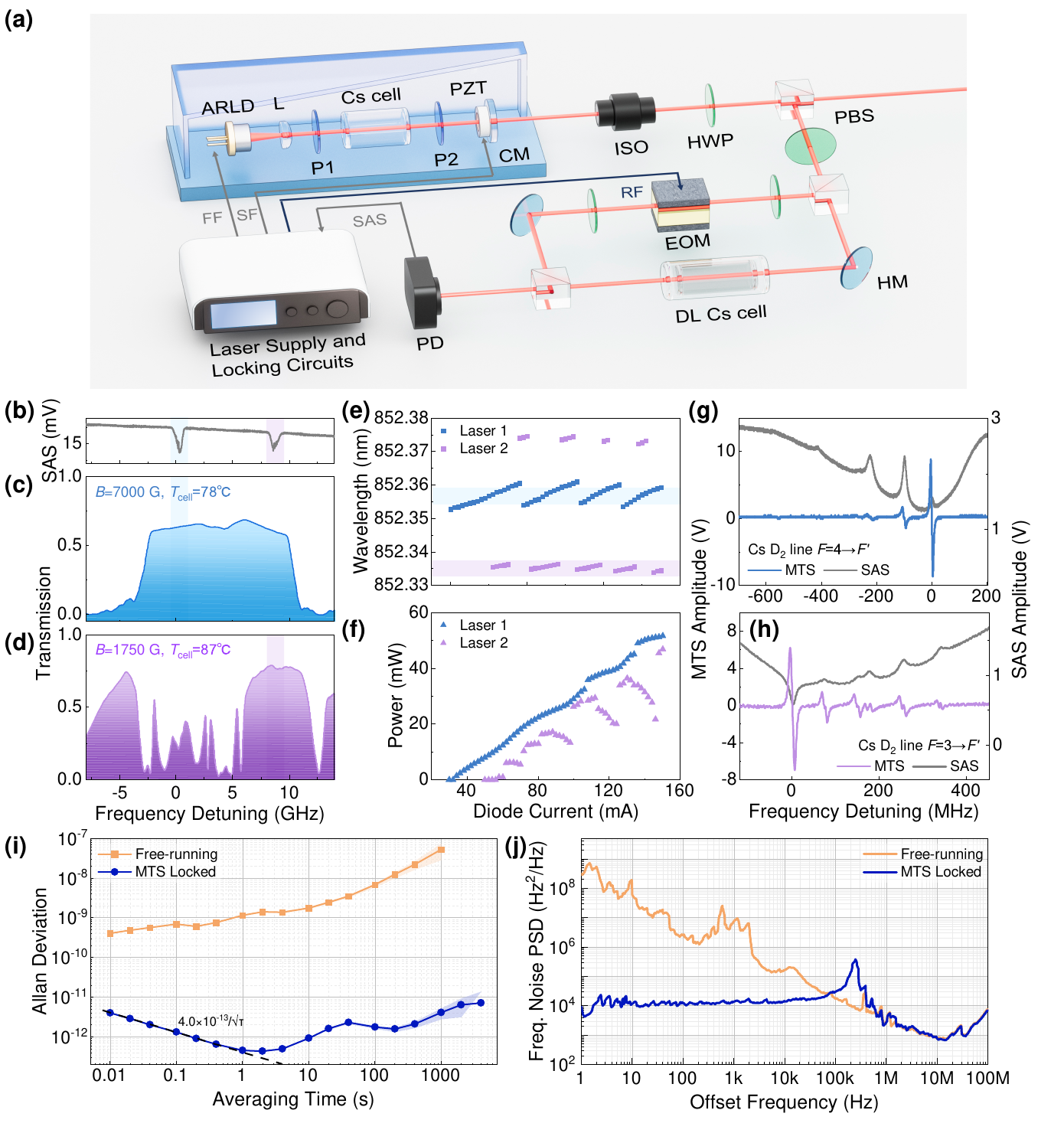}
\caption{Characteristics of the atom-referenced Faraday laser scheme. (a)~Schematic of the atom-referenced Faraday laser with MTS locking. FADOF transmission spectra for Laser~1~(c) and Laser~2~(d) under optimized working parameters, with the (b)~SAS signal for reference. The zero-detuning frequency corresponds to the Cs D$_2$ line $F=4 \rightarrow F'=5$ transition. (e)~Wavelength and (f)~output power versus diode current for the free-running Laser~1 (blue data) and Laser~2 (purple data), respectively. The scanned MTS error signal and SAS reference signal for Laser~1~(g) and Laser~2~(h), where the zero-detuning frequencies correspond to their cycling transitions, respectively. (i)~Allan deviation of the heterodyne beat between the two lasers under free-running (orange) and MTS-locked (blue) conditions. (j)~Frequency noise PSD of a single laser before (orange) and after (blue) MTS locking, with extracted Lorentzian linewidths of 2.32 and 2.12~kHz, respectively. ARLD, antireflection-coated laser diode; L, lens; P1 and P2, linear polarizers; PZT, piezoelectric transducer; CM, cavity mirror; ISO, isolator; EOM, electro-optic modulator; DL Cs cell, double-layer Cs cell; FF, fast feedback to ARLD; SF, slow feedback to PZT; RF, radio-frequency modulation frequency to EOM; MTS, modulation transfer spectroscopy; SAS, saturated absorption spectroscopy.}
\label{Fig.2}
\end{figure}

\subsection{Atom-referenced Faraday Lasers}

We demonstrate two atom-referenced Faraday lasers as the narrow-linewidth and low-noise optical sources for the DFP Cs clock. The experimental setup of the Faraday laser and its MTS locking system is shown in Fig.~\ref{Fig.2}(a). Light emitted by the antireflection-coated laser diode (ARLD) is collimated by an aspheric lens (L) and then passes through the FADOF for frequency selection. The FADOF consists of a 10-mm-long Cs vapor cell placed between two crossed linear polarizers, P1 and P2. Under appropriate cell-temperature and axial magnetic-field conditions, the polarization of light at frequencies near the Cs transitions is rotated by approximately $90^\circ$ as it propagates through the cell, allowing the light to pass through P2. Under the external cavity feedback provided by the partially reflecting cavity mirror (CM) with a reflectivity of 40\%, the laser oscillation is established. The external-cavity length is about 30~mm, corresponding to a free spectral range (FSR) of 5~GHz. To suppress laser-frequency fluctuations arising from mechanical- and temperature-induced cavity-length variations, the cavity structure is machined from a single piece of low-expansion Invar. The CM is bonded directly to the Invar base without an adjustable mount, thereby improving the mechanical and thermal stability of the cavity. A piezoelectric transducer (PZT) attached to the CM is used to tune the external-cavity length. The temperature control of ARLD and Cs vapor cell is realized using a thermoelectric cooler (TEC) and a film heater, respectively, together with a thermistor and proportional-integral-derivative (PID) controller with a precision of 1~mK.

The two pumping transitions impose different FADOF-frequency-selection requirements. Laser~1 targets the $F=4 \rightarrow F'=5$ transition of the Cs D$_2$ line. To achieve robust laser operation over a wide tuning range with stable output power, a strong magnetic field of 7000~G is applied and the cell temperature is maintained at 78~\textcelsius. Under these conditions, the Cs atoms enter the hyperfine Paschen-Back regime and exhibit pronounced Zeeman-sublevel splitting, generating a relatively broad, flat-top FADOF transmission peak covering the $F=4 \rightarrow F'$ and $F=3 \rightarrow F'$ transitions, with a full width at half maximum (FWHM) of approximately 13~GHz, as shown in Fig.~\ref{Fig.2}(c). Together with the 5-GHz external-cavity FSR and mode competition, the transmission spectrum enables stable single-mode operation near the $F=4 \rightarrow F'$ transitions. The corresponding wavelength and power characteristics are shown in Fig.~\ref{Fig.2}(e) and (f), respectively. The laser can be tuned from 852.3528~nm to 852.3610~nm and reaches a maximum output power of 51.7~mW.

For laser~2 aiming at the $F=3 \rightarrow F'=2$ transition, a lower magnetic field is required to maintain operation in the anomalous Zeeman regime, producing a transmission peak centered near $F=3 \rightarrow F'$ while effectively suppressing transmission corresponding to the $F=4 \rightarrow F'$ transitions. Based on the theoretical analysis of the FADOF transmission spectrum under high optical intensity, an optimized transmission spectrum with a FWHM of 5.8~GHz is obtained at a magnetic field of 1750~G and a cell temperature of 87~\textcelsius, as shown in Fig.~\ref{Fig.2}(d). This configuration enables stable single-mode operation over the wavelength range from 852.3339~nm to 852.3364~nm, with a maximum output power of 46.8~mW. In addition, a second wavelength range is observed near 852.37~nm, which is associated with the additional transmission peak at a detuning of approximately -5~GHz in Fig.~\ref{Fig.2}(d). However, it does not affect operation at the $F=3 \rightarrow F'=2$ transition when the laser is stabilized by MTS. The intrinsic alignment of the FADOF transmission peaks with the atomic transitions contributes to the turnkey operation and long-term frequency robustness of the laser system.

\subsection{MTS-locked Faraday Lasers}
\label{mts}
To further suppress laser frequency noise and improve frequency stability for the performance enhancement of DFP clock, both Laser~1 and Laser~2 are stabilized to their respective Cs D$_2$ line cycling transitions via MTS. Unlike Pound--Drever--Hall (PDH)~\cite{2025-SB-Loong}, MTS does not require a high-finesse reference cavity, strict mode matching, or complex cavity-length stabilization. It is therefore well suited to compact and deployable systems~\cite{2024-nature-sea,2025-nc-sea,2025-PR-lzj}. Compared with saturated absorption spectroscopy (SAS)~\cite{2001-josab-sas}, MTS provides a nearly Doppler-background-free signal with a steeper zero-crossing error signal slope, thereby improving the locking performance.

As shown in Fig.~\ref{Fig.2}(a), the laser entering MTS locking system is divided into a strong pump beam and a weak probe beam. The pump beam is phase-modulated at 4.95~MHz by an electro-optic modulator (EOM) and counterpropagates with the probe beam through a double-layer Cs vapor cell. Based on the nonlinear four-wave-mixing process of MTS, the modulation is transferred from the pump to the probe beam. The latter is subsequently detected by a photodetector (PD), and the resulting electrical signal is demodulated, filtered and amplified to generate a dispersive error signal. The error signal is then processed by a PID controller and divided into fast and slow feedback branches applied to the ARLD and PZT, respectively. The locking system incorporates automated spectral-search and lock-acquisition routines. After an occasional loss of lock, the system typically reacquires the target transition and restores the lock within 30~s, thereby supporting robust long-term operation of the Cs clock. The corresponding SAS and MTS signals of Laser~1 and Laser~2 are shown in Fig.~\ref{Fig.2}(g) and (h), where the zero-detuning frequencies are defined with respect to the $F=4 \rightarrow F'=5$ and $F=3 \rightarrow F'=2$ cycling transitions for the two lasers, respectively.

The frequency stability of the lasers before and after MTS locking is evaluated via the optical heterodyne beat-note measurement between Laser~1 and Laser~2. As shown in Fig.~\ref{Fig.2}(i), the fractional frequency stability, characterized by the beat-note Allan deviation (ADEV) normalized to the optical carrier frequency, follows $4.0 \times 10^{-13}/\sqrt{\tau}$ between $\tau=$0.01~s and 1~s under MTS locking. The long-term frequency stability at $\tau=$ 1,000~s reaches $4.1\times10^{-12}$, which is approximately four orders of magnitude better than the free-running result. The factors that influence the MTS-locking performance include laser power fluctuations, residual amplitude modulation (RAM) induced by the etalon effect and the EOM crystal birefringence effect, temperature drift of vapor cell, EOM and analog servo electronics, etc. Future work will address these limitations to further enhance the long-term frequency stability.

Furthermore, we characterize the frequency-noise PSD (obtained from the phase-noise measurement) of a single laser and estimate its Lorentzian linewidth. In this measurement, Laser~2 is tuned to the same Cs D$_2$ $F=4\rightarrow F'=5$ transition as Laser~1 by adjusting the FADOF temperature, and an AOM introduces a frequency offset for heterodyne detection. Assuming that the two lasers contribute equal and statistically independent frequency noise, $S_{\nu,\mathrm{beat}}(f)\approx 2S_{\nu,\mathrm{single}}(f)$, we divide the beat-note PSD by two to estimate the single-laser PSD, as shown in Fig.~\ref{Fig.2}(j). The free-running atom-referenced laser already exhibits a low white-frequency-noise floor. MTS locking further suppresses frequency noise below 75~kHz, while introducing a servo-related peak near 245~kHz. The noise floors are $738~\mathrm{Hz}^2/\mathrm{Hz}$ in the free-running state and $675~\mathrm{Hz}^2/\mathrm{Hz}$ under MTS locking, corresponding to the Lorentzian linewidths of 2.32 and 2.12~kHz, respectively, according to $\Delta\nu_{\mathrm{L}}=\pi S_{\nu}^{\mathrm{white}}$~\cite{2019-APLP-lorentz}. Thus, the atom-referenced cavity provides a low intrinsic white-noise floor, whereas MTS locking suppresses the low-frequency noise for the stable and high-SNR DFP Cs clock operation.

\begin{figure}[htbp]
\centering
\includegraphics[width=10 cm]{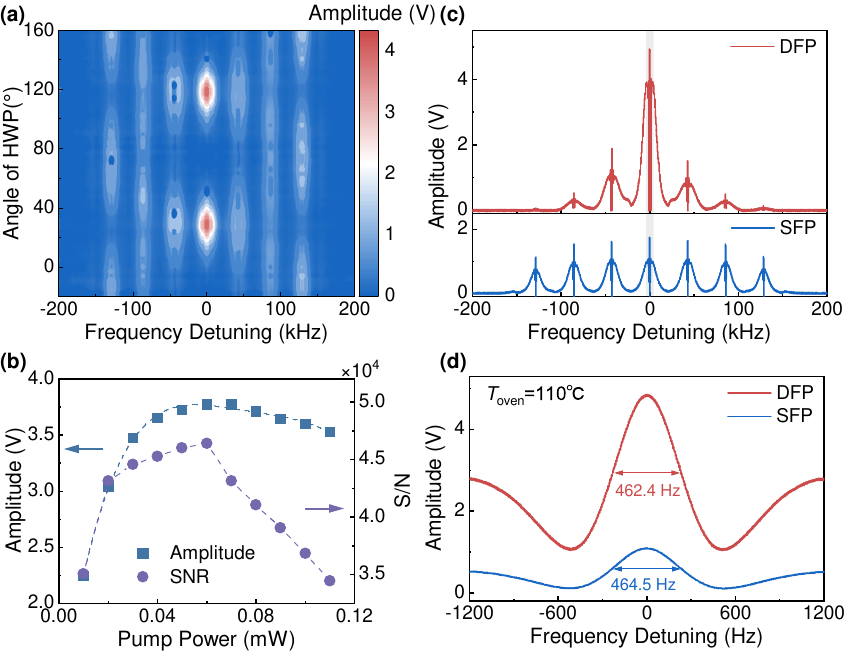}
\caption{Ramsey pattern and SNR optimization of DFP Cs beam clock. (a) When the Cs oven temperature $T_{\mathrm{oven}}=$ 110~\textcelsius, the Ramsey pattern changes with the polarization of Pump~2. The polarization is determined by the HWP angle. (b) Clock signal amplitude (blue squares) and SNR measured at 137~Hz with a 1-Hz bandwidth (purple circles) as functions of Pump~2 power. (c) Ramsey pattern under DFP (red curve) and SFP (blue curve). (d) The clock transition Ramsey signal of DFP (red curve) and SFP (blue curve) clock. The zero detuning corresponds to 9,192,631,770~Hz.}
\label{Fig.3}
\end{figure}

\section{RESULTS AND DISCUSSION}
\subsection{SNR Optimization}
\label{S3.1}

To determine whether the increased atomic utilization provided by DFP can be converted into an SNR improvement, we optimize the polarization and power of pump and probe lasers. Here we present the optimization process of Pump~2, which directly determines the population accumulated in the $|F=3,\mathrm{m_F}=0\rangle$ clock state. All measurements are performed at $T_{\mathrm{oven}}=110$~\textcelsius, with the probe and Pump~1 powers fixed at their optimized values of 0.9 and 0.8~mW, respectively. By rotating the HWP angle determining the polarization of Pump~2, we characterize the atomic population distribution among the seven Ramsey resonance peaks, as shown in Fig.~\ref{Fig.3}(a). Initially, at an HWP angle of $-20^\circ$, the highest peak of the Ramsey pattern corresponds to $m_\mathrm{F}=\pm 3$. As the HWP angle increases, the peak gradually shifts to $m_\mathrm{F}=0$. Notably, at HWP angles of $27^\circ$ and $117^\circ$, the polarization state is closest to $\pi$ polarization, with the $m_F=0$ peak reaching its maximum, while the $m_F=\pm 3$ peaks are substantially suppressed. The residual populations in the $m_F=\pm1$ and $\pm2$ states may originate from imperfect polarization of pump lasers due to the residual magnetic-field inhomogeneity in the pumping region. As the Cs tube is fully vacuum-sealed, direct measurement of the actual magnetic field in the pumping region is difficult.

Furthermore, with the HWP angle fixed at $27^\circ$, we examine the effect of Pump~2 power on the amplitude and SNR of the clock signal, as shown in Fig.~\ref{Fig.3}(b). The signal amplitude reaches its maximum at 0.07~mW and subsequently decreases because stronger Pump~2 excitation transfers an increasing fraction of atoms back to \(F=4\). The maximum SNR of 46,365, however, occurs at the slightly lower power of 0.06~mW. This difference reflects the simultaneous increase in clock-signal noise with optical power. The clock-signal noise is measured using a Fast Fourier Transform (FFT) network analyzer (SRS SR770) at 137~Hz with a 1-Hz bandwidth.

The direct comparison between the DFP and single-Faraday-laser-pumping (SFP) configurations is shown in Figs.~\ref{Fig.3}(c) and (d). Under SFP operation, the atomic population is approximately distributed among the seven Zeeman sublevels, and the $0\rightarrow0$ clock transition accounts for only about 14\% of the total Ramsey-pattern amplitude. Introducing Pump~2 increases this fraction to 52\%. Consequently, the clock-signal amplitude increases by a factor of 3.8, from 0.99 to 3.77~V. Although the noise also increases from 45.4 to $81.31~\mu\mathrm{V}/\sqrt{\mathrm{Hz}}$, the signal enhancement dominates, producing a net 2.1-fold increase in SNR from 21,806 to 46,365 in a 1-Hz bandwidth. Meanwhile, the Ramsey linewidth remains essentially unchanged, which is 464.5 and 462.4~Hz under SFP and DFP operation, respectively. These results demonstrate that the enhanced clock-state population provided by DFP is converted into a genuine SNR improvement rather than merely a larger Ramsey signal.

\begin{figure}[t]
\centering
\includegraphics[width=8cm]{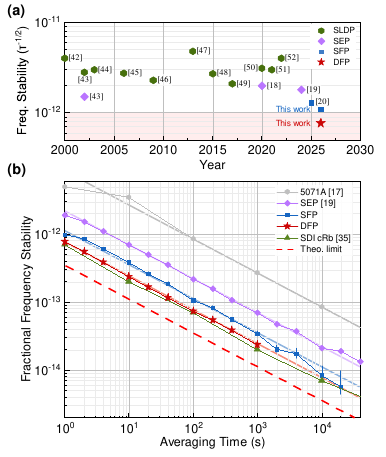}
\caption{(a) Summary of reported compact Cs clock frequency stabilities with different laser-pumping architectures, including single-laser-diode pumping (SLDP, green hexagons)~\cite{2000-UFFC-M,2002-IFCS-M,2003-IFCS-M,2006-EFTF-M,2009-FSM-M,2016-EFTF-M,2015-ITSF-M,2017-IFCS-M,2020-RSI-M,2021-AO-hexuan,2022-FT-M}, single-ECDL pumping (SEP, purple diamonds)~\cite{2002-IFCS-M,2020-OE-SHS,2024-PR-STT}, single-Faraday-laser-pumping (SFP, blue squares)~\cite{2025-prapplied-shb}, as well as the DFP (red stars) scheme demonstrated in this work. (b) The frequency stability of Cs beam clocks under different pumping schemes, theoretical stability calculated under DFP (red dotted line), the performance of commercial Cs beam clock 5071A~\cite{5071A} and commercial cold Rb atomic clock SDI cRb-clock~\cite{crb-clock}.}
\label{Fig.4}
\end{figure}

\subsection{Short-term Frequency Instability}

We evaluate the fractional frequency stability of the DFP Cs clock by comparing its 10~MHz output against an active hydrogen maser (VCH-1003M Option L), as shown in Fig.~\ref{Fig.4}(b). Benefiting from the improved SNR, the short-term stability of the DFP Cs clock reaches $7.7 \times 10^{-13}/\sqrt{\tau}$, breaking through the conventional $10^{-12}/\sqrt{\tau}$ stability level of compact Cs beam clocks. This performance is better than that of the SFP clock under comparable experimental conditions using only Laser~1, with a stability of $1.1 \times 10^{-12}/\sqrt{\tau}$. And it is about one-order-of-magnitude better than the commercial Microchip 5071A clock\cite{5071A}. Remarkably, the stability reached by the DFP Cs clock at an averaging time of $10^3$~s is already comparable to the $10^4$~s stability previously achieved in an SEP Cs beam clock based on IF-ECDL~\cite{2024-PR-STT}. These results validate that our atom-referenced laser scheme successfully suppresses the technical frequency noise that traditionally limits two-laser-pumped Cs clocks, enabling the compact Cs clock to enter the $10^{-13}/\sqrt{\tau}$ stability regime and approach state-of-the-art medium-term stability within a shorter averaging time.

Based on the measured SNR of 46,365 and Ramsey linewidth of $\Delta \nu=462.4$~Hz, the detection-noise-limited short-term stability of DFP Cs clock is estimated as 
\begin{equation}
\sigma_y(\tau)=\frac{1}{\pi}\frac{\Delta \nu}{\nu_0} \frac{1}{ \mathrm{SNR}}\frac{1}{\sqrt{\tau}},
\end{equation}
where $\nu_0$ represents the ground-state hyperfine transition frequency of the $^{133}$Cs atom. This estimate gives a short-term stability of $3.45\times10^{-13}/\sqrt{\tau}$. The measured instability is approximately 2.2 times higher than this detection-noise-limited value. It shows that, although the optical SNR enhancement improves the clock output, the present system does not yet operate at its SNR-predicted limit, indicating the presence of additional technical noise.

Possible contributions include the short-term stability of the oven-controlled crystal oscillator (OCXO), additive phase noise introduced by the microwave-synthesis chain, and noise from the clock-frequency servo and its associated electronics. The present microwave chain is referenced to the 5-V low-noise option of a Morion MV197 OCXO, whose specified 1-s ADEV is below $1\times10^{-12}$, which is of the same order as the measured short-term clock stability. In the future, replacing the present OCXO with a lower-noise oscillator and further optimizing the microwave-synthesis and servo electronics are expected to provide a measured stability closer to the SNR-estimated result.

To summarize how laser architecture and linewidth impact the performance of compact optically pumped Cs beam clocks, we listed the reported short-term frequency stabilities over the past two decades, as shown in Fig.~\ref{Fig.4}(a). Conventional single-laser-diode pumping (SLDP) schemes based on DFB/DBR laser sources, typically characterized by MHz-level linewidths~\cite{2000-UFFC-M,2002-IFCS-M,2003-IFCS-M,2006-EFTF-M,2009-FSM-M,2016-EFTF-M,2015-ITSF-M,2017-IFCS-M,2020-RSI-M,2021-AO-hexuan,2022-FT-M}, generally lead to limited frequency stabilities above $2\times10^{-12}/\sqrt{\tau}$. In comparison, single-ECDL-pumped (SEP) architectures benefit from reduced linewidth and improved spectral purity, enabling consistently better stability~\cite{2002-IFCS-M,2020-OE-SHS,2024-PR-STT}. Further improvement is achieved in SFP schemes, owing to the atom-referenced frequency selection and a narrower laser linewidth of 2.3~kHz~\cite{2025-prapplied-shb}. The present DFP clock reaches $7.7\times10^{-13}/\sqrt{\tau}$, bringing the short-term fractional frequency stability of a compact Cs beam clock into the $10^{-13}/\sqrt{\tau}$ regime. These comparisons indicate that combining low-noise atom-referenced lasers with dual pumping provides a practical route beyond the conventional low-$10^{-12}/\sqrt{\tau}$ stability level. Furthermore, theoretical calculations predict that introducing hexapole magnetic focusing (HMF) in the optical-pumping region could increase the effective atomic utilization by a factor of 9.5 through transverse confinement of the Cs atomic beam~\cite{2022-FiP-hex}. Combined with the DFP architecture demonstrated here, this enhancement is expected to improve the short-term fractional frequency stability of compact Cs beam clock to below $3\times10^{-13}/\sqrt{\tau}$.

\subsection{Long-term Frequency Instability}

\begin{figure}[t]
\centering
\includegraphics[width=12 cm]{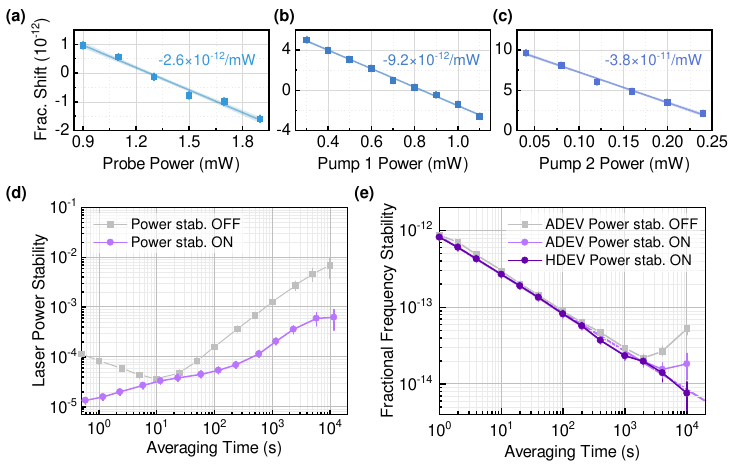}
\caption{(a)--(c) Linear-fit slopes of fractional frequency shift versus the probe, Pump~1 and Pump~2 power are $-2.6\times 10^{-12}/\mathrm{mW}$, $-9.2\times 10^{-12}/\mathrm{mW}$ and $-3.8\times 10^{-11}/\mathrm{mW}$, respectively. (d) The relative power stability of Laser~1 measured before (grey squares) and after (purple circles) laser power stabilization. (e) The long-term frequency stability of DFP Cs beam clock evaluated from Allan deviation (ADEV) and Hadamard deviation (HDEV).}
\label{Fig.5}
\end{figure}

For the medium- and long-term stability measurement, we observe a drift trend in the ADEV at averaging time above 2,000~s, as shown in Fig.~\ref{Fig.5}(e) (grey data). We consider the main contribution is optical-power-dependent frequency shifts. Residual pump and probe light reaching the Ramsey region through scattering or reflection would introduce the AC Stark shift, while atomic fluorescence and power-dependent variations in the Ramsey signal and clock discriminator may also introduce additional shifts. To evaluate the power sensitivities of DFP Cs clock, we monitor the fractional frequency shift of the signal compared between DFP clock and hydrogen maser while independently varying the probe, Pump~1, and Pump~2 powers. The linear fitting coefficients near the nominal operating points yield $k_{\mathrm{probe}}=-2.6\times10^{-12}\mathrm{mW}^{-1}$, $k_{\mathrm{Pump1}}=-9.2\times10^{-12}\mathrm{mW}^{-1}$ and $k_{\mathrm{Pump2}}=-3.8\times10^{-11}\ \mathrm{mW}^{-1}$, respectively, as shown in Fig.~\ref{Fig.5}(a)--(c). For the laser with relative power stability of $6.8\times 10^{-3}$ at 10,000~s, the power-dependent DFP clock frequency stability is approximately $5.5\times10^{-14}$, which is close to the measured result of $5.3\times10^{-14}$ shown in Fig.~\ref{Fig.5}(e) (grey data). 

To optimize the long-term stability, active power stabilization is implemented using a liquid-crystal variable retarder (LCVR) for Laser 1 (including probe and Pump~1 laser beams) and the +151.2~MHz-frequency-shift AOM for Pump~2. Each power stabilization system is controlled by a digital PID loop. As shown in Fig.~\ref{Fig.5}(d), the fractional power stability of Laser 1 is effectively improved after LCVR power stabilization, and the ADEV at 10,000~s is reduced from $6.8\times10^{-3}$ to $6\times10^{-4}$. The power stability of Laser 2 exhibits a comparable result for $\tau>1,000$~s. Meanwhile, the clock ADEV follows approximately $8.2\times10^{-13}/\sqrt{\tau}$ from 1~s to 4,000~s, and its value at 10,000~s is reduced from $5.3\times10^{-14}$ to $1.8\times10^{-14}$, as shown in Fig.~\ref{Fig.5}(e). The corresponding HDEV is $7.4\times10^{-15}$, which is insensitive to linear frequency drift, providing a complementary measure of the clock stability in the presence of residual slow drift. 

The residual stability drift contributions at long averaging times are considered to be slow laser locking frequency drift and microwave-power-dependent frequency shifts. As discussed in Sec.~\ref{mts}, the MTS locking effect is influenced by RAM and thermal drift of the MTS system. In the future, three approaches will be pursued to improve the long-term clock stability. First, the FADOF transmission profile will be optimized to provide a flatter operating region, thereby improving the optical-power stability. Second, the optical and electronic sources of MTS locking frequency drift will be suppressed. Expanding the laser beams will improve intensity uniformity, while wedged EOM and vapor-cell windows and active EOM temperature control will suppress parasitic etalon effects and RAM. The present analog locking circuits will be replaced with digital servo controllers incorporating low-temperature-coefficient components, stable voltage references, and improved thermal isolation. Third, active stabilization of the microwave power delivered to the Ramsey cavity, together with improved microwave shielding and thermal stabilization of the synthesis chain, should further suppress microwave-induced frequency shifts.

\section{CONCLUSION}

In this work, we demonstrate a compact DFP Cs beam clock based on two atom-referenced Faraday lasers. The combination of FADOF-based atom-referenced external-cavity feedback and MTS stabilization enables turnkey laser operation with a Lorentzian linewidth of 2.12~kHz, thereby reducing laser-frequency noise and the associated FM-to-AM noise conversion in the clock signal. With the optimized $F=4\rightarrow F'=4$ $\sigma$-polarized and $F=3\rightarrow F'=3$ $\pi$-polarized pumping configuration, the clock-state atomic utilization improves to 52\%. The Ramsey-fringe SNR reaches 46,365 in a 1-Hz bandwidth, improving the short-term fractional frequency stability from $1.1\times10^{-12}/\sqrt{\tau}$ under SFP operation to $7.7\times10^{-13}/\sqrt{\tau}$ under DFP operation. Active optical-power stabilization further suppresses slow power-dependent frequency shifts, yielding an HDEV of $7.4\times10^{-15}$ at $10^4$~s. These results demonstrate a practical approach for translating the enhanced atomic utilization of two-laser pumping into improved clock stability, providing a promising route toward high-performance, field-deployable frequency standards. In the future, combining sufficient suppression of residual technical noise with hexapole magnetic focusing, which is predicted to increase effective atomic utilization 9.5-fold~\cite{2022-FiP-hex}, could enable the DFP clock to achieve a short-term fractional frequency stability better than $3\times10^{-13}/\sqrt{\tau}$.

\begin{backmatter}

\bmsection{Funding}
National Natural Science Foundation of China (62405007); China Postdoctoral Science Foundation (BX2021020); Quantum Science and Technology-National Science and Technology Major Project (2021ZD0303200); Hebei Provincial Natural Science Foundation Basic Research Special Project - 2025 Basic Research Program Proof-of-Concept Project (F2025109009).

\bmsection{Disclosures}
The authors declare no conflicts of interest.

\bmsection{Data availability} Data underlying the results presented in this paper are not publicly available at this time but may be obtained from the authors upon reasonable request.

\end{backmatter}

\bibliography{sample}






\end{document}